\documentstyle[multicol,aps,epsf,rotate]{revtex}

\begin{document}
\input psfig.tex

\title{Scaling of Energy Barriers for Flux Lines and Other Random Systems}
\author{Barbara Drossel and Mehran Kardar}
\address{Department of Physics, Massachusetts Institute of
Technology, Cambridge, Massachusetts 02139}
\date{\today}
\maketitle
\begin{abstract}
Using a combination of analytic arguments and numerical simulations,
we determine lower and upper bounds for the energy barriers to
the motion of a defect line in a random potential
at low temperatures. We study the cases
of magnetic flux lines in high-$T_{c}$ superconductors
in 2 and 3 dimensions,
and of domain walls in 2 dimensional random-field Ising models.
The results show that, under fairly general conditions, energy
 barriers have the same scaling as the fluctuations in free energy,
except for possible logarithmic factors.
This holds not only for barriers between optimal configurations of
the line, but also for
barriers separating any metastable configuration from a configuration
of minimal energy. Similar arguments may be applicable to other
elastic media with impurities, such as bunches of flux lines.

\end{abstract}

\begin{multicols}{2}

\section{Introduction}
\label{u1}

Directed paths in random media (DPRM) \cite{KardarRev} are
 simple realizations of glassy systems \cite{glass1,glass2}.
Some examples are pinned flux lines (FL) in high-$T_c$ superconductors,
and domain walls in random field and random bond Ising models.
In thermal equilibrium,
 a magnetic FL is pinned by defects (oxygen impurities, grain
boundaries, etc.) in the superconductor which lower
its energy \cite{expreview}. The resulting elastic distortions are
 limited by the line tension which
opposes the bending of the line. This comptetition leads to a free energy
landscape for the FL which is rather complicated and has many local
minima, i.e. metastable states \cite{BoseGlass}.
When an electric current flows
through the system, the FL feels a Lorentz force perpendicular to its
orientation and to the current direction.
As long as the current is not strong enough to overcome the pinning
forces, the line moves by thermally
activated jumps of line segments between metastable configurations
\cite{KimAnderson,FisherFisherHuse,JoffeVinokur}.
The length of these line segments is estimated by the condition
that the free energy barrier for a jump should be of the same
order as the gain in free energy due to that jump. These dynamics
are believed to be the reason for the nonlinear
voltage-current characteristics found in experiments \cite{expreview}.

Randomly placed impurities in Ising ferromagnets may generate either a random
magnetic field or random exchange couplings \cite{HuseHenley}.
The free energy landscape for domain walls in these systems
is determined by the competition between the pinning energy and
the energy cost (per unit length or area) for creating the wall.
When the system is quenched to a low temperature, the magnetic domains
grow. As for the flux line, the free energy gain due to the motion of a
domain wall segment is expected to be of the same order as the free energy
barrier
which has to be overcome.

Since energy barriers play an important role in the dynamics
of glassy systems, it is essential to know their properties. The scale
of these barriers should grow with observation size $L$ like
a power law $L^\psi$. Usually, it is assumed that the only energy
scale in the system is set by the fluctuations in free
energy which increase as
$L^\theta$, and that therefore $\psi = \theta$
\cite{HuseHenley,JoffeVinokur}. However,
it is also quite possible that the heights of the ridges in
the random energy landscape scale differently from those of the
valleys  that  they separate, with $\psi>\theta$.  Yet another
scenario is that transport occurs mainly along a percolating
channel of exceptionally low energy valleys with $\psi<\theta$.

A first attempt to clarify this situation was taken in Ref.\cite{MDK},
 where $\psi = \theta$ was established for a FL moving in 2 dimensions.
Using a combination of analytic arguments and numerical
simulations, lower and upper bounds to the barrier were found.
This argument was then extended to a FL in 3 dimensions \cite{barrier3d},
yielding again $\psi = \theta$.
In this article, we present in more detail the arguments discussed briefly in
these earlier papers,  including also systems with
long-range correlated randomness (random-field Ising models) in 2 dimensions.
We obtain in all cases lower and upper bounds to the barrier that scale
as $L^\theta$, except for possible logarithmic factors,
leading to $\psi = \theta$. Furthermore, it is argued that the result
$\psi = \theta$ holds also in higher dimensions, as long as the distribution
of minimal energies decays exponentially. In all cases, the line can
move through the system by encountering energy fluctations of only order
$L^\theta$ around the mean minimal energy. We also show that a line which
initially has a larger energy can reach this region of minimal energies
by crossing barriers of order $L^\theta$ (or smaller).

The outline of the paper is as follows: In section \ref{u2}, we determine
the energy barrier for a FL moving in 2 dimensions. In section \ref{u3}, we
apply the same algorithm to determine the energy barrier to the motion of
 domain walls in 2-dimensional random-field Ising systems. In
section \ref{u4}, we study energy barriers for a FL in 3 dimensions and discuss
also the behavior in higher dimensions. In section \ref{u5}, we take a
 general look at the energy landscape and show that a line can
move from any initial configuration to a minimal configuration by
going over no barrier higher than $L^\theta$. In section \ref{u6}, we
try to put the definition of energy barriers on a more solid foundation,
and section \ref{u7} argues that the results of the paper can be
generalized to other elastic media with impurities.

\section{Energy barriers for flux lines in 2 dimensions}
\label{u2}

In two dimensions, we represent a DPRM by the following model:
The line is  discretized
to lie on the bonds of a square lattice, directed along its
diagonal. Each segment of the line can proceed along
one of two directions, leading to a total of $2^t$ configurations
after $t$
steps. These configurations are labelled by the set of integers $
\left\{ x(\tau)
\right\}$ for $\tau=0,1,\cdots,t$, giving the transverse coordinate
of the line
at each step (clearly constrained such that $x(\tau+1)=x(\tau)\pm
1$). To each
bond on the lattice is assigned a (quenched) random energy equally
distributed
between 0 and 1. The energy of each configuration is the sum of all
random bond energies on the line. Without loss of generality, we set
$x(0)=0$.

\begin{figure}
\narrowtext
\centerline{{\epsfysize=3in
\epsffile{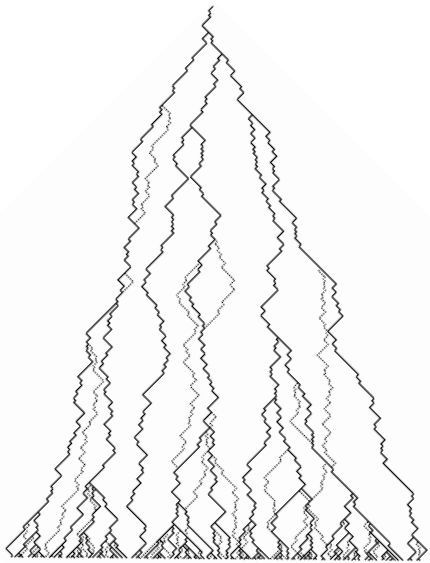}}}
\caption{
Minimal paths of length $t = 256$ for a FL in 2 dimensions
to endpoints between
$x = -96$ and  $x = +96$ (solid),
and the barrier paths between them (dotted).
}
\label{MinimalPaths1}
\end{figure}

Some exact results are known for this model \cite{KardarRev}:
The fluctuations in the free energy at finite
temperature scale as $t^{1/3}$. The meanderings of the transverse coordinate
of the line scale as $t^\zeta$, where $\zeta = 2/3$ is the
roughness exponent. The scaling behavior of the
pinned FL is governed by a zero-temperature fixed point
\cite{HuseHenley} where energy fluctuations scale in the same way.
A FL at low temperatures, and in thermal equilibrium, is likely to
spend most of the time in configurations of minimal energy.
For each endpoint $(t,x)$ with $x = -t, -t + 2, \cdots, t$, there is
a configuration of minimal
energy $E_{min}(x|t)$ which can be obtained numerically in a time of
order $t^2$.
It is  known that for $|x| < x_c \propto t^{2/3}$, the function
$E_{min}(x|t)$ behaves as a random walk and is thus asymptotically Gaussian
distributed \cite{KardarRev,HuseHenleyFisher}.
Since beyond the interval $[-x_c,x_c]$ the energy of minimal paths is
systematically larger,
 we consider in this paper only the region $[-x_c,x_c]$.
Fig.\ref{MinimalPaths1} shows minimal paths of length
$ t = 256 $ to endpoints between $x = -96$ and $x = +96$.

We want to find the energy  barrier that
has to be overcome when  the line is moved from an initial minimal
energy
configuration $\{x_i(\tau)\}$  between $(0,0)$ and $(t,-x_f)$ to a final
configuration $\{x_f(\tau)\}$   between
$(0,0)$ and $(t, +x_f)$, with $x_f \equiv x_f(t) \le x_c$.
The only elementary move allowed is flipping
a kink along the line from one side to the other (except at the end point).
Thus the point  $(\tau, x)$ can
be shifted to $(\tau, x \pm 2)$. Each route from the initial to the
final configuration is
obtained by a sequence of such elementary moves. For each sequence,
there
is an intermediate configuration of maximum energy, and a barrier
which
is the difference between this maximum and the initial energy. In a
system at
temperature  $T$, the probability that the FL  chooses a sequence
which crosses a barrier of height $E_B$ is proportional to
$\exp(-E_B/T)$, multiplied by the number of such sequences. We assume
that, as is the case for the equilibrium DPRM, the ``entropic''
factor of the number of paths does not modify scaling behavior. Thus
at sufficiently low temperatures, the FL chooses the optimal sequence
which has to overcome the least energy, and the overall barrier is
the minimum
of barrier energies of all sequences.

Since the number of elementary moves
scales roughly as the area between the initial and final lines, the
number of possible
sequences grows as $t^{xt}$. This exponential growth makes
it practically impossible to find the barrier by examining all
possible sequences,
hampering a systematic examination of barrier energies. Rather than
finding
the true barrier energy, we proceed by placing upper and lower bounds
on it.

\begin{figure}
\narrowtext
\centerline{\rotate[r]{\epsfysize=3in
\epsffile{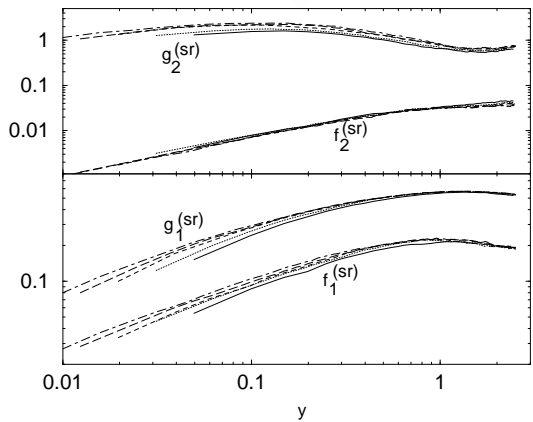}}}
\vskip 1cm
\caption{Scaling functions $f_1^{(sr)}(y)$, $f_2^{(sr)}(y)$,
$g_1^{(sr)}(y)$, and $g_2^{(sr)}(y)$
(see Eqs.(1), (4),
and (5)) for the mean and variance of the
lower and upper bounds; averaged over 2000 realizations of
randomness,
for $t = 256$ (solid), $512$ (dotted), $1024$ (dashed),
$2048$ (long dashed), and $ 4096$ (dot--dashed).
}
\label{Barriers1}
\end{figure}

The lower bound was given in  Ref.\cite{HwaFisher}, and is obtained
as follows:
The endpoint  of the path has to visit all sites $(t,x)$ with $|x|
\leq x_f$, and  the
energy of  any path ending at $(t,x)$ is at least as large as
$E_{min}(x|t)$. Therefore the barrier
energy cannot be smaller than $\max[E_{min}(x|t) - E_{min}(-x_f|t)]$
for ${x \in [-x_f, x_f]}$.
When $x_f$ is sufficiently small, the probability distribution of
this lower bound
is identical to that of the maximal deviation of a random
walk of length $x_f$ \cite{Mikheev}.
The latter is a Gaussian distribution with a  mean value proportional to
$\sqrt{x_f}$, and a variance scaling as $x_f$. This growth
saturates for
$x_f$ of the order of $t^{2/3}$, leading to the scaling behaviors,
\begin{eqnarray}
  \left\langle {E_-^{(sr)}} (t,x) \right\rangle  & = & t^{1/3} f_1^{(sr)}(x /
t^{2/3}),\qquad \text{ and}
\nonumber \\
 {\rm var}(E_-^{(sr)})& = & t^{2/3} f_2^{(sr)}(x / t^{2/3}) ,
\end{eqnarray}
for the lower bound and its variance.
The functions $f_1^{(sr)}(y)$ and $f_2^{(sr)}(y)$ are proportional to
$\sqrt{y}$
and $y$ for small $y$, respectively,  and go to a
constant for $y = O(1)$. Our simulation results for systems with $t=$
256, 512,
1024, 2048, and 4096 confirm this expectation. Fig.\ref{Barriers1} shows
the scaling functions $f_1^{(sr)}(y)$ and $f_2^{(sr)}(y)$ for different $t$,
and
the collapse is quite
satisfactory. However, the initial growth  proportional to $ \sqrt{x_f}$, is
not clearly seen at these sizes.

To obtain an upper bound for the barrier, we specify an explicit
algorithm for
moving the line from its initial to final configuration. This is
achieved by
finding a sequence of intermediate steps. It is certainly
advantageous to keep the
intermediate paths as close to minimal configurations as possible. We
therefore proceed in the following way:
We first find the minimal paths connecting $(0,0)$ to the points
$(t - 1, x)$ with $x_i(t - 1) < x < x_f(t - 1)$, and we add a
last step to the left (from $(t - 1, x)$ to $(t, x - 1)$).
If $x_i(t - 1) > x_i(t) = -x_f$, we then move the point $(t, -x_f)$
to $(t, -x_f + 2)$. Now the path has the same endpoint as the first
intermediate minimal path. We then  move the path to this first
intermediate configuration (the precise prescription will be
given below), and then we move again the endpoint. This procedure
is repeated, until the path reaches its final configuration.
At each step, we obtain a local barrier path which separates two
neighboring minimal
configurations. The overall barrier is of course the one with the
highest energy.
While it may occasionally be possible to go to the next intermediate
configuration in a single elementary move (as defined above), this
is
generally not the case. Intermediate minimal paths with the same
endpoint may  be
quite far apart  at coordinates $\tau < t$. The reason is simple:
suppose the
random potential has a large positive fluctuation, a ``mountain.''
The minimal
energy paths will then circumvent this region by going to its right
or left. The last
path going to the left and the first one going to the right have
almost the same energy.
They form a loop which  can be quite large and  is likely to enclose the
barrier when
both paths separate already at small $\tau$. Such loops have been
conjectured \cite{JoffeVinokur,FisherFisherHuse} to play an important
role
 in the low-temperature dynamics of the DPRM. Since the transverse
fluctuations of a minimal path of length $t$ grow as $t^{2/3}$, we
expect the
lateral size of these loops to also be of this order.

The algorithm for moving a line of length $t = 2^n$ from an intermediate
configuration $\{x_1(\tau)\}$  to another
one $\{x_2(\tau)\}$, with $x_2(t) = x_1(t)$ is as follows:
If $x_2(\tau) \le x_1(\tau) + 2$ for all $\tau$, we can choose a
sequence of
elementary moves such that at most two bonds of the line are not on
one
or the other minimal path, leading to a barrier of order 1 between
the two.
If $x_2(\tau) > x_1(\tau) + 2$ for some $\tau$, the two paths enclose
a loop. We then
consider the points ${(t / 2 - 1, x)}$ with $x_1(t / 2 - 1) < x <
x_2(t / 2 - 1)$.  For each of
these points, we find a minimal segment of length $t / 2 - 1$, connecting
the point   ${(t / 2 - 1, x)}$ to   $(0,0)$ by a minimal path,
and we take a final step to the left from
$(t / 2 - 1, x)$ to $(t / 2, x - 1)$.
  In the same way, we connect the
points ${(t / 2 + 1, x)}$ with $x_1(t / 2 + 1) < x <
x_2(t / 2 + 1) $ to $x_1(t))$ via minimal paths
and add a first step to the right from $(t / 2, x - 1)$ to  $(t / 2, x)$.
Two such segments form together an almost minimal path of length $t$,
constrained to go through intermediate points at $t / 2$
and $t/2 \pm 1$.  We next move the line  $\{x_1(\tau)\}$
stepwise through
this sequence of almost minimal paths.  If $x_i(t / 2 - 1) = x_i(t / 2) - 1$,
we first move the upper segment. If  $x_i(t / 2) = x_i(t / 2 + 1) + 1$,
we then move the lower segment. Then we move the middlepoint. We
continue by repeatedly moving the upper segment, the lower segment, and the
middle point, until the final configuration  $\{x_2(\tau)\}$ is reached.
(If the length of the line is different from $2^n$,
we might have to choose the upper segment to be larger
by 1 than the lower segment, or vice versa.)

The prescription for moving the segments of length $t / 2$  is
exactly the same as for
paths of length $t$: If the distance between two consecutive
configurations is larger than
2 for some  $\tau \in [0, t / 2]$, we consider the points at ${(t /
4 \pm 1, x)}$  in between the
two, and find minimal paths of length $t / 4 - 1$ connecting them to the
initial and final points, and add a step to the middle points.
Next we attempt to move segments of length $t / 2$ by
repeatedly moving
line portions of length $t / 4$. In some cases, when the energy
barrier is large, it
is necessary to proceed with this construction until the cutoff scale  of
$t / 2^{n - 1} = 2$ is
reached. Thus, at each intermediate configuration, the line is
composed of one
segment of length $t/2$, one of length $t/4$, etc; ending
with two smallest pieces
of  length $t / 2^m$ (equal to 2 in the worst case). The barrier
path is the intermediate configuration with highest energy.
Fig.\ref{MinimalPaths1} shows the barrier paths resulting from the above
construction.

We now estimate the barrier energy resulting from the above
construction.
Each intermediate path is composed of segments of minimal paths with
constrained
endpoints, and we would like to find the probability distribution for
the highest energy.
Constraining the endpoints of a minimal path of length $\tau$
typically increases
its energy by $E_-^{(sr)}(\tau)\propto \tau^{1/3}$. A subset of these
intermediate paths
(those that cross the largest mountains) have constraints imposed on
segments
of length $t$, $t/2$, $t/4$, and all the way down to unity.  The
number of paths in this
subset (henceforth referred to as  candidate barriers) grows as
$N_c(t)\propto t^\alpha$, with $1<\alpha<1+2/3$.
The lower limit comes from noting that for each loop of size $2^m$ there
exist at least two loops of size $2^{m-1}$,
 one in the upper and one in the
lower half of the parent loop, thus $N_c\geq t$.
The upper limit comes from the
total number of intermediate configurations
 that grows as $tx_f$. The energy of each candidate
barrier path is obtained in a manner similar to that of the lower bound:
Instead of finding
the maximum of a random walk of length $x_f\propto t^{2/3}$, we now
have to examine
the sum of the maxima for a sequence of shorter and shorter random
walks added
together. The mean value of this sum is related to the convergent
series,
\begin{eqnarray}\label{meanEc}
&&\left\langle E_c^{(sr)}(t) \right\rangle=\nonumber\\
&& = \left\langle
E_-^{(sr)}(t)+ 2\, E_-^{(sr)}(t/2)+ 2\, E_-^{(sr)}(t/4)+\cdots \right\rangle
+A\ln(t) \nonumber\\
&&=\left\langle
E_-^{(sr)}(t) \right\rangle\left(1 + 2\, (2^{-1/3} + 2^{-2/3} +
\cdots)\right) +A \ln(t) + B
\nonumber \\
&& \simeq \left\langle E_-^{(sr)}(t)\right\rangle
\left(-1 + 2(1-2^{-{1/3}})^{-1} \right)
+ A\ln(t) + B \nonumber\\
&&= 8.69... \left\langle E_-^{(sr)}(t) \right\rangle+A\ln(t) + B.
\end{eqnarray}
The correction term $A\ln(t)$, is explained as follows:
Each segment of length $2^m$ is composed of a minimal path of length
$2^m - 1$ and one step which has a random energy (the final or initial step,
depending on whether the segment lies in the upper or lower half of a loop).
So the energy of the segment is equal to the energy $E_-^{(sr)}(2^m)$
of a minimal path of length $2^m$, plus a constant of order 1. Since a
candidate barrier has $n = \ln(t) / \ln(2)$ segments, these constants
add up to $A \ln(t)$, with $A$ of the order of unity.
The constant $B$ in Eq.(\ref{meanEc}) accounts for the
breakdown of the scaling form of the energy increase  for small
loops.
The mean angle of the smallest loops (of size 2)
approaches the $45^\circ$
limit; their mean energy growing as $0.5 t_m$. For the larger loops,
the angle
$t_m^{2/3} / t_m$ is small and the energy is $0.23 t_m$. A finite
value of $m$
acts as a cutoff separating the two limits.  The energy difference
per unit length between
small and large paths then leads to the additive constant $B$ (of the
order of unity)
in Eq.(\ref{meanEc}).

The barrier energy is the maximum of the $N_c(t)$ energies of all
candidate
barriers. To find its characteristics, we need the whole probability
distribution for
the energy $E_c^{(sr)}(t)$. Since $E_c^{(sr)}$ is the sum of energies
coming from  its minimal
segments, the simplest assumption is to regard the segment
energies as
independent, approximately Gaussian, random variables. We then
conclude that $E_c^{(sr)}(t)$ is also Gaussian distributed with a variance,
\begin{eqnarray}\label{varEc}
&& {\rm var}\left( E_c^{(sr)}(t) \right) \nonumber\\
&&={\rm var}\left( E_-^{(sr)}(t)\right)+
2\,\left({\rm
var}\left( E_-^{(sr)}(t/2)\right)+
\cdots\right) \nonumber\\
&&\simeq 4.40\ldots {\rm var}\left( E_-^{(sr)}(t) \right) \propto t^{2/3}.
\end{eqnarray}
Since the different segments are in fact constructed through a
specific recursive
procedure, their independence cannot be justified. In the worst
case that they are completely dependent, the right-hand side of
Eq.(\ref{varEc}) has to be multiplied by $n = \log_2(t)$. Since our
numerical results show no evidence for such a logarithmic factor, we shall not
consider it any further.

It can be checked easily that (for large $N$), the maximum of $N$
independent Gaussian variables of mean $a$ and variance $\sigma^2$, is a
Gaussian of mean
$a+\sigma\sqrt{2\ln N}$ and variance $\sigma^2/(2\ln N)$ \cite{Galambos}.
Since the  $N_c(t)$
candidate barriers have large segments in common, their energies are
not
independent. We can approximately take this into account by
assuming a subset of them as independent, leading to  $N\propto
t^{\alpha'}$ for
some $\alpha' < \alpha$. We thus obtain the following estimates for
the mean upper
bound in barrier energy,
\begin{eqnarray}\label{meanE+1}
\left\langle E_+^{(sr)}(x,t) \right\rangle&=& \left\langle E_c^{(sr)}(x,t)
\right\rangle
+\sqrt{2\ln N{\rm var} E_c^{(sr)}(x,t)}\nonumber\\
&\simeq& \left( 1 +\beta^{(sr)} \sqrt{\ln t} \right)t^{1/3} g_1^{(sr)}(x /
t^{2/3}),
\end{eqnarray}
and its variance,
\begin{eqnarray}\label{varE+}
{\rm var}\left( E_+^{(sr)}(x,t) \right)&=&{{\rm var}\left( E_c^{(sr)}(x,t)
\right)\over 2\ln N^{(sr)}} \nonumber\\
&\simeq& {t^{2/3} \over \ln t}g_2^{(sr)}(x / t^{2/3}).
\end{eqnarray}
The functions $g_1^{(sr)}(y)$ and $g_2^{(sr)}(y)$ are proportional to
$\sqrt{y}$
and $y$ respectively, for small $y$,
constant at large $y$, and in general different from those of the
lower bound.

Our numerical simulations indeed confirm the above scaling forms. The
scaling
functions $g_1^{(sr)}(y)$ and $g_2^{(sr)}(y)$ are plotted in
Fig.\ref{Barriers1} for
different values
of $t$, after averaging over 2000 realizations of randomness. The
$\sqrt{\ln(t)}$
factors are essential, as a comparable collapse is not obtained
without them.
In fact, the best fit to $<E_+^{(sr)}(t)>$ is obtained by including the
correction to scaling
term $ \propto <E_-^{(sr)}(t)>$,
and with $\beta^{(sr)} = 1$. The numerics
therefore  support the
neglect of correlations, and the assumption of a Gaussian distributed
$E_c^{(sr)}(t)$.
As in the lower bound, the initial scaling proportional to $\sqrt{ x_f}$ is
not clearly seen
for the sizes studied.  Since the leading power for the scaling of
the lower and upper
bounds is identical, we conclude that the barrier energies also grow
as $t^{1/3}$.
(It remains to be seen if the logarithmic factors are truly
present, or merely an
artifact of our algorithm.)

\section{Energy barriers for domain walls in 2-dimensional
 random field Ising systems}
\label{u3}

In the previous section, we considered random bond
energies which were uncorrelated. The analytic
argument for the upper bound relied on the random-walk behavior
of $E_{min}(x|t)$ in
this situation. Thus, the proof for $\psi
= \theta$ can not directly be extended to other situations,
where the distribution of lower bound energies is not known.
An important example is the case of domain walls in 2-dimensional
random-field Ising magnets. The energy for creating a domain
wall is equal to the cost of flipping all spins
on one side of the interface; in turn proportional to the sum of
all random fields on the flipped spins. There are consequently
long-range correlations in the domain wall energy in the direction
perpendicular to the orientation of the wall\cite{MKJAP}.

We describe the configurations
of the domain wall by essentially the same model as the FL, but assigning
to each bond a random energy with long-range correlations in the
$x$-direction.
These correlations are generated by first selecting for each time $t$, random
numbers $\{r_t^{(-N)}, r_t^{(-N+1)}, \cdots , r_t^{(N-3)},r_t^{(N-1)}\}$
equally
distributed between $-1$ and $1$, where $N$ is (at
least) as large as the largest time occuring in the simulations. To each
bond connecting $(t,x)$ to $(t + 1, x\pm 1)$ we then assign the energy
\begin{displaymath}
{1\over \sqrt{2N}}
\left(\sum_{i = -N}^{x - 1/2 \pm 1/2} r_t^i - \sum_{x + 1/2 \pm 1/2}^{N-1}
r_t^i \right) \, .
\end{displaymath}

\begin{figure}
\narrowtext
\centerline{{\epsfysize=3.5in
\epsffile{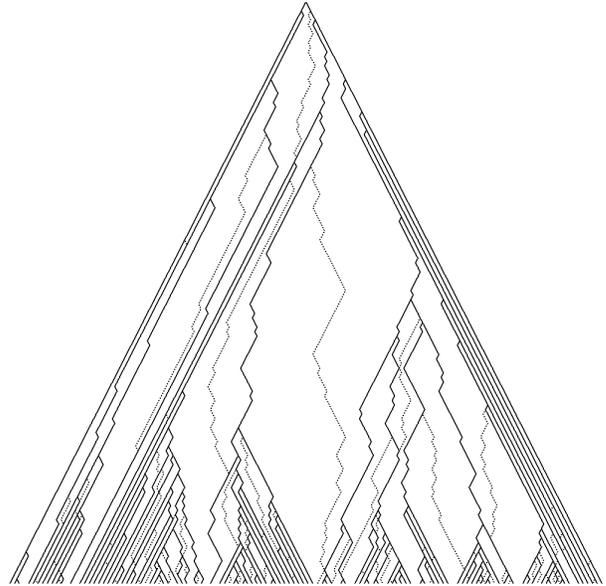}}}
\caption{
Minimal paths of length $t = 128$ to endpoints between
$x = -t$ and  $x = +t$ (solid),
and the barrier paths between them (dotted), in the random-field
Ising model.
}
\label{MinimalPaths2}
\end{figure}

\begin{figure}
\narrowtext
\centerline{\rotate[r]
{\epsfysize=3.5in
\epsffile{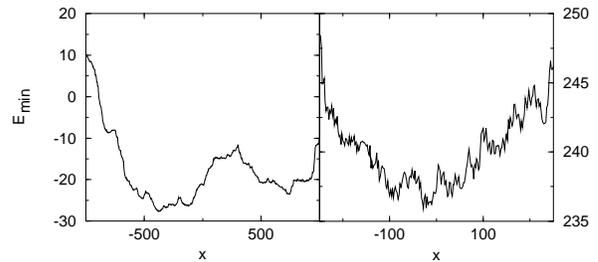}}}
\caption{
The minimal energy $E_{min}$ as function of the endpoint position $x$
for $t = 1024$ and (a) long-range correlated randomness, (b) short-range
correlated randomness.
}
\label{walk}
\end{figure}

Fig.\ref{MinimalPaths2} shows minimal paths of length
$ t = 128 $. Due to the correlations, neighboring bonds have almost
the same energy, and therefore minimal paths tend to have large
parallel portions.
Fig.\ref{walk} shows the minimal energy as function of the endpoint
position for a given realization of randomness, and for $t = 1024$.
This curve is much smoother and has longer correlations
than the corresponding curve in the case of short-range correlated
randomness, where the minimal energy performs a random walk.

\begin{figure}
\narrowtext
\centerline{\rotate[r]
{\epsfysize=3.5in
\epsffile{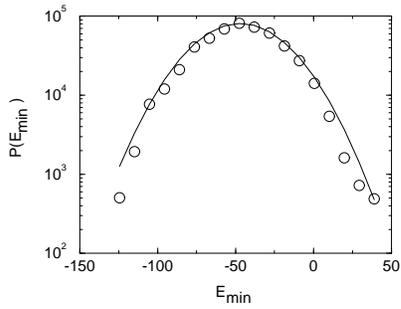}}}
\caption{
The distribution $P(E_{min})$ of minimal energies for $t = 1024$ and
long-range correlated randomness. The solid line is a Gaussian distribution;
deviations from it indicate a third cumulant.
}
\label{Emin2}
\end{figure}

The fluctuations in free energy of a line are known to scale as
$t$, and the roughness exponent is $\zeta = 1$\cite{correlated}.
We determined numerically the
distribution function for the minimal energy shown in Fig.\ref{Emin2}.
It is very close to a Gaussian, with no apparent power-law tails.
We will show that, due to this property of the minimal energy distribution,
the lower and upper bounds to the barrier scale in the same way as
the fluctuations in minimal energy.

\begin{figure}
\narrowtext
\centerline{\rotate[r]{\epsfysize=3.5in
\epsffile{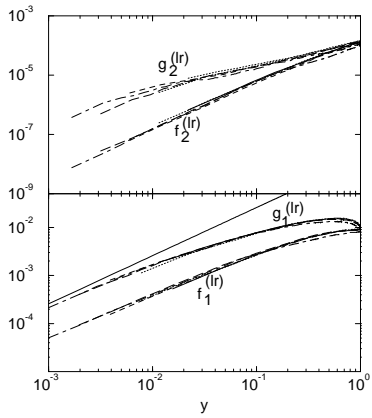}}}
\caption{Scaling functions $f_1^{(lr)}(y)$, $f_2^{(lr)}(y)$,
$g_1^{(lr)}(y)$, and $g_2^{(lr)}(y)$
(see Eqs.(6), (8),
and (9)) for the mean and variance of the
lower and upper bounds; averaged over 1000 realizations of
randomness,
for $t = 256$ (solid), $t512$ (dotted), $1024$ (dashed), and
$2048$ (long dashed). The straight line has the slope 1.
}
\label{Barriers2}
\end{figure}

\begin{figure}
\centerline{\rotate[r]{\epsfysize=3.5in
\epsffile{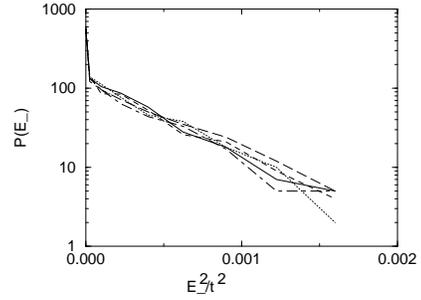}}}
\narrowtext
\caption{ Distribution of minimal energy for an Ising domain wall
in 2 dimensions
with random fields. The symbols are the same as
in the previous figure.
}
\label{distrlower}
\end{figure}

As in the previous section, we move the line from an initial minimal
energy
configuration $\{x_i(\tau)\}$  between $(0,0)$ and $(t,-x_f)$, to a final
configuration $\{x_f(\tau)\}$   between
$(0,0)$ and $(t, x_f)$.
Since the endpoint  of the path has to visit all sites $(t,x)$ with $|x|
\leq x_f$, and since the
energy of  any path ending at $(t,x)$ is at least as large as
$E_{min}(x|t)$, the barrier
energy cannot be smaller than $\max[E_{min}(x|t) - E_{min}(-x_f|t)]$.
Since the distribution of minimal energies
decays exponentially and has no power-law tails, we can expect that the
lower bound scales in the same way as the fluctuation of the minimal
energy, leading to
\begin{eqnarray}
  \left\langle {E_-^{(lr)}} (t,x) \right\rangle  & = & t f_1^{(lr)}(x / t),
\qquad \text{ and}
\nonumber \\
 {\rm var}(E_-^{(lr)})& = & t^2 f_2^{(lr)}(x / t) ,
\end{eqnarray}
for the lower bound and its variance.
 Our simulation results for systems with $t=$
256, 512,
1024 and 2048 confirm this expectation. Fig.\ref{Barriers2} shows
the scaling functions $f_1^{(lr)}(y)$ and $f_2^{(lr)}(y)$ for different $t$,
and  the collapse is quite
satisfactory. However, the initial growth  proportional to $ {x_f}$, is
not clearly seen at these sizes.
Fig.\ref{distrlower} shows the distribution of lower bound energies.
It is very close to a (half)-Gaussian of width proportional to $ t$.

An upper bound can be obtained by exactly the same algorithm as before.
The analytic argument made in the previous section, however,
cannot be directly repeated, since the function $E_{min}(x|t)$ is no longer
a random walk in $x$, and since we do not have analytic results for the lower
bound. We can, however, combine analytic arguments with the numerical
results for the lower bound to predict the scaling behavior of the upper
bound. Since the line is always composed of minimal segments, the energy
of a candidate barrier which has segments of all lengths down to the cutoff
is given by
\begin{eqnarray}\label{meanEc2}
&&\left\langle E_c^{(lr)}(t) \right\rangle \nonumber\\
&&=\left\langle
E_-^{(lr)}(t)+2\,\left(E_-^{(lr)}(t/2)+E_-^{(lr)}(t/4)+\cdots\right)
 \right\rangle \nonumber\\
&&\simeq  3 \left\langle E_-^{(lr)}(t) \right\rangle + A' \ln(t) + B'.
\end{eqnarray}
The origin of the terms $A' \ln(t) + B'$ has been
 explained in the previous section
(see paragraph after Eq.(\ref{meanEc})).
 Since the energy distribution of the lower bound is
approximately (half-)Gaussian, the energy distribution of
the candidate barriers decays also like a Gaussian.
The upper bound to the barrier energy is the maximum of the energies
of all candidate barriers. In our simulations, we find no evidence
for logarithmic factors, indicating that the number of candidate
barriers increases either very slowly, or not at all, with $t$.
{}From
Fig.\ref{MinimalPaths2} we can see that there is essentially one large
loop over a distance of the order of the length of the path, leading
to only few independent candidate barriers.
As in the
previous section, we find that the maximum of these  candidate barriers,
which is the upper bound to the barrier energy,
scales in the same way as the lower bound, i.e.
\begin{eqnarray}\label{meanE+2}
\left\langle E_+^{(lr)}(x,t) \right\rangle&=& \left\langle E_c^{(lr)}(x,t)
\right\rangle
+\sqrt{2\ln N^{(lr)}{\rm var} E_c^{(lr)}(x,t)}\nonumber\\
&\simeq& t\, g_1^{(lr)}(x / t),
\end{eqnarray}
and its variance scales as
\begin{eqnarray}\label{varE+2}
{\rm var}\left( E_+^{(lr)}(x,t) \right)&=&{{\rm var}\left( E_c^{(lr)}(x,t)
\right)/ 2\ln N^{(lr)}} \nonumber\\
&\simeq& {t^{2}} g_2^{(lr)}(x / t).
\end{eqnarray}
Fig.\ref{Barriers2} shows the scaling functions $f_2^{(lr)}$
and $g_2^{(lr)}$.

To summarize the results so far, we have established the relation
$\psi = \theta$
for lines in 2-dimensional systems with  short- and long-range correlated
randomness. Since the considerations for
both systems rely strongly on the dimensionality, it is of importance
to look also at a 3-dimensional system, which is physically more relevant.

\section{Energy barriers for a flux line in 3 dimensions}
\label{u4}

In a two-dimensional system, the endpoint of the FL has to move through
all points $(x,t)$ with $x_i < x < x_f$. This property was essential for
the derivation of the lower bound in the previous sections.
A FL which moves in three
dimensions can avoid regions in space which are energetically unfavorable
for some of its segments, and one may therefore
speculate that $\psi<\theta$.
In this section, we first determine numerically a lower bound for the
barrier
energy which scales in the same way as the energy fluctuations,
thus ruling out $\psi<\theta$. Further numerical results
 predict that an upper bound scales in the same way,
thus leading to $\psi = \theta$.

The line now lies on the bonds of a cubic lattice, starting
at the origin and  directed
along its (1,1,1) diagonal. Each segment of the line can proceed
in the positive direction along
one of the three axes, leading to a total of $3^t$
configurations  after $t$ steps, with endpoints lying in the plane
which is
spanned by the points $(t,0,0)$, $(0,t,0)$, and $(0,0,t)$.
 A given configuration of the FL is labelled
 by vectors $\left\{ \vec x(\tau)
\right\}$ for $\tau=0,1,\cdots,t$, giving the transverse coordinates
of the FL at each step. The
points  $\left\{ \vec x(\tau)\right\}$ lie on the
vertices of a triangular lattice. For a given value of $\tau$, they
lie on one of three alternating sublattices.

\begin{figure}
\centerline{{\epsfysize=3.5in
\epsffile{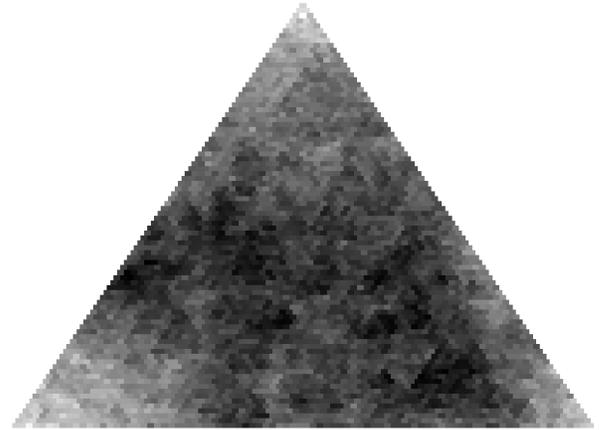}}}
\vskip -3true cm
\narrowtext
\caption{ Minimal energies of paths of length $t = 288$ in 3 dimensions
to endpoints $\vec x$ with $|\vec x| < O(t^\zeta)$. The higher energies
are indicated by white shading, and the lower energies by darker shades.
}
\label{profile}
\end{figure}

\begin{figure}
\narrowtext
\centerline{\rotate[r]{\epsfysize=3.5in
\epsffile{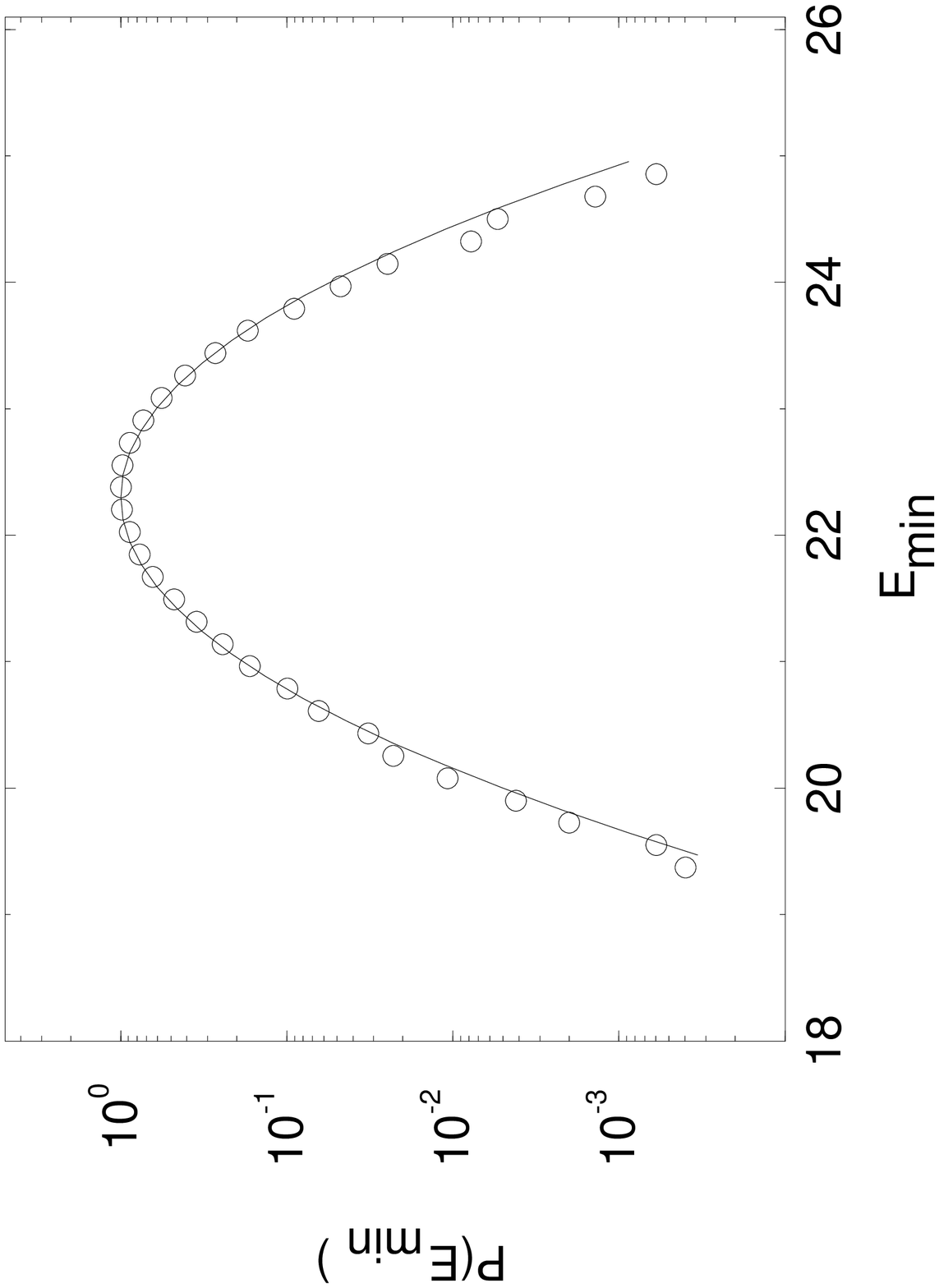}}}
\vskip 1true cm
\caption{Probability distribution $P(E_{min})$
of minimal energies $E_{min}(\vec 0|144)$ in 3 dimensions,
averaged over 50,000 realizations of randomness.
The solid line is a Gaussian
distribution.
}
\label{Emin3}
\end{figure}

The minimal energy $E_{min}(\vec x |t)$ can be obtained
numerically in a time of order $t^3$.
The fluctuations in minimal
energy are known to scale as $t^\theta$ with $\theta \simeq 0.24$,
and the roughness exponent for minimal paths
is $\zeta \simeq 0.62$ \cite{AmarFamily,KimBrayMoore}.
The endpoints of the minimal paths with the lowest energy lie
within a distance proportional to $ t^\zeta$ from the origin. Figure
\ref{profile}
shows the minimal energies of paths of length $t = 288$ to
endpoints $\vec x$ with $|\vec x| < O(t^\zeta)$. The highest energy in
this region is represented in white, the smallest energy in black.
The minimal energies are correlated over
a distance of the order of $t^\zeta$. The distribution of minimal
energies is close to a Gaussian and is shown in Fig.\ref{Emin3}.
Similar to a 2-dimensional system \cite{HalpinHealy}
(see also Fig.\ref{Emin2}),
this distribution seems to have a third cumulant since it is not
completely symmetric.

A lower bound to the barrier energy is obtained as follows:
While the line moves from its initial to final
configuration, the transverse coordinates of its endpoint move
 between nearest-neighbor
positions on one of the above mentioned triangular sublattices.
When the endpoint is at a position $\vec x$,
the energy of the line is at least as large as the minimal energy
$E_{min}(\vec x |t)$. The maximum of
all these minimal energies along the trajectory of the endpoint,
minus the energy of the initial configuration, certainly bounds
the barrier energy from below. Since we do not know the actual
trajectory of the endpoint, we have to look for the trajectory
with the smallest maximal energy. Only in this way
can we be sure that we have indeed found a lower bound. This
situation is fundamentally different from a 2-dimensional system,
where there is only one possible trajectory for the endpoint.

\begin{figure}
\narrowtext
\centerline{\rotate[r]{\epsfysize=3.5in
\epsffile{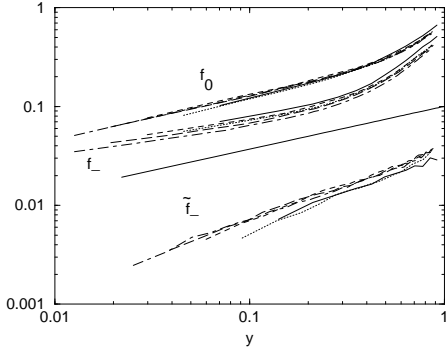}}}
\caption{Scaling functions $f_-(y)$, $\tilde f_-(y)$, and $f_0(y)$
defined in Eqs. (10)--(12)
for $t = 72$ (solid), $144$ (dotted), $288$ (dashed),
$ 576$ (long dashed), and $1152$ (dot-dashed),
averaged over 500 realizations of randomness. The straight line has
slope $\theta/\zeta = 0.39$.
}
\label{lower}
\end{figure}

\begin{figure}
\narrowtext
\centerline{\rotate[r]{\epsfysize=3.5in
\epsffile{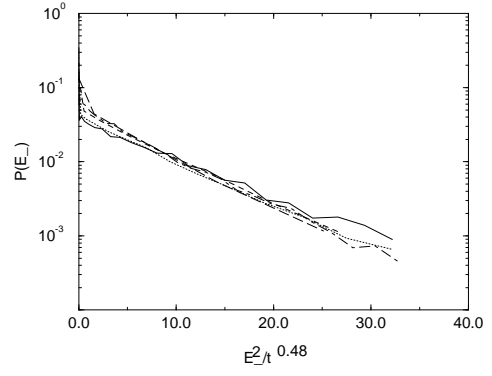}}}
\caption{
Probability distribution of $E_-$. The parameters
and symbols are the same as in Fig.10.
}
\label{distribution}
\end{figure}

Provided that the minimal energies $E_{min}(\vec x |t)$ are known,
this lower bound is determined
in polynomial time by using a transfer-matrix method:
We start by assigning to the initial point $\vec x_i$ a ``barrier
energy'' $B(\vec x_i) = 0$, and to all other sites $\vec x$ on the same
sublattice a barrier energy $B(\vec x) = t$,
which is certainly larger than the lower bound resulting
 from the algorithm after many iterations. At each step, the
energy $B$ of all sites $\vec x$, except for the initial site, is updated
according to the following rule: Look for the minimum of the energies
 $B(\vec x \pm \vec e_i)$ of the 6 neighbors. If this is smaller than
 $B(\vec x)$, replace  $B(\vec x)$
by this minimum, or by $E_{min}(\vec x|t) - E_{min}(\vec x_i|t)$,
whichever is larger. After a sufficiently large number of
iterations,
which is of the order of the size of the area of interest
(scaling as $t^{2\zeta}$), all possible trajectories to endpoints
within this area have been probed, and the barrier energies $B(\vec x)$ do not
change any more. The energy  $B(\vec x_f)$ is then identified
as the lower bound. Figure \ref{lower} shows the lower bound to the
energy barrier for a line with the endpoint moving from the origin
to sites within a distance of the order of
$t^\zeta$, for different values of $t$ and averaged over 500
realizations of randomness. The distance $|\vec x_f - \vec x_i|$ has been
scaled by $t^{\zeta}$, and the energy by $t^{\theta}$. After
this rescaling, all the curves collapse, leading to
the following scaling behavior for the lower bound,
\begin{equation}
\left\langle {E_-} (t,|\vec x_f - \vec x_i|) \right\rangle   =
 t^{\theta} f_-(|\vec x_f - \vec x_i| / t^{\zeta}).\label{eq1}
\end{equation}
The function $f(y)$ should be proportional to $y^{\theta/\zeta}$
for small $y$. Again, for the simulated system sizes,
this asymptotic scaling is not clearly seen.
For $y > 1$, the scaling form in Eq.(\ref{eq1}) breaks down
since the minimal energy is then a function of the angle $(|\vec x| / t)$.
We conclude that the lower bound to the barrier scales in the same way
as the
fluctuations in minimal energy, and consequently the energy
barrier increases at least as $t^\theta$, leading to $\psi \ge
 \theta$.
The distribution $P(E_-)$ of the lower bound energy
for a fixed distance $|\vec x| \propto t^\zeta$ is
 shown in Fig.\ref{distribution}.
It appears to be half-Gaussian with width proportional to $t^\zeta$.

The result $\psi \ge  \theta$ is not surprising if we note
that an even simpler lower bound is given by $\max(E_{min}(\vec x_f|t)
- E_{min}(\vec x_i|t), 0)$, which evidently
scales as $t^\theta$ since the
distribution function of minimal energies decays exponentially
fast, i.e. has no power-law tails (see Fig.\ref{Emin3}).
To make sure that the scaling of the lower bound found above is not
dominated by the neighborhood of final configurations with particularly
high energies, we repeated the above simulations
 by allowing only endpoints with minimal
energies smaller than the initial energy.
This corresponds to a situation where the endpoint of the line only
moves to positions which are energetically more favorable.
The results are
shown in Fig.\ref{lower} and again collapsed by the scaling form
\begin{equation}
\left\langle {\tilde E_-} (t,|\vec x_f - \vec x_i|) \right\rangle   =
 t^{\theta} \tilde f_-(|\vec x_f - \vec x_i| / t^{\zeta}). \label{eq2}
\end{equation}
As in the previous case, the asymptotic scaling $\tilde f_-(y) \propto
y^{\theta/\zeta}$ for small $y$ cannot be clearly seen. The energy
distribution of the lower bound is again a half-Gaussian of width
proportional to $ t^{\zeta}$ and looks similar
to Fig.\ref{distribution}.

The same scaling behavior is also found when instead of the
optimal trajectory for the endpoint, the shortest trajectory
(a straight line) is chosen. In this case, the mean of the barrier
energy $E_0$ has the scaling form
\begin{equation}
\left\langle {\tilde E_0} (t,|\vec x_f - \vec x_i|) \right\rangle
 = t^{\theta} f_0( |\vec x_f - \vec x_i|/ t^{\zeta})\label{eq3}
\end{equation}
(see Fig.\ref{lower}), again with a half-Gaussian
distribution of width proportional to $t^\zeta$.
This, of course, does not represent a lower bound to the true
barrier, but it will be important for the determination of
an upper bound below, and is therefore included here.

The result $\tilde E_- \propto t^\theta$ (Eq.(\ref{eq2}))
can be explained from the exponential
tails of the distribution of minimal energies:
If we asume that the
endpoint of the line moves only in valleys of particularly low
energy, we can successively remove all sites with the largest
 minimal energy from the set of possible endpoints, until the connectivity
over the distance $t^\zeta$ breaks down.
The remaining endpoints form
 percolation clusters, and their density is given by the
corresponding percolation threshold. (This is analogous to random resistor
networks describing the hopping resistivity for strongly localized electrons.
The resistance of the whole sample is governed by the critical resistor
that makes the network percolate \cite{randomresistor}.) Since the occupied
sites are
correlated over the distances considered, the value for the
threshold is different from the site percolation limit of 0.5 in
an infinite triangular lattice with no correlation between occupied
sites. But for the present purpose, it is sufficient to know that
this threshold is finite, and that therefore a finite percentage
of all sites are below  threshold. Since the distribution of
minimal energies decays rapidly, its tail cannot contain a finite
percentage of all sites. We conclude that the threshold is
within a distance of $t^\theta$ from the peak, and therefore that
 the energy fluctuation on the percolation cluster, and
consequently the lower bound for the barrier, are proportional to
$t^\theta$.

We now proceed to construct an upper bound to the energy
barrier.  To this purpose, we specify a sequence of
elementary moves which take the line from its initial to
final configuration.
The only elementary move allowed is flipping a
kink along the line. Thus the point $(\tau, \vec x)$ can
be shifted to $(\tau, \vec x \pm \vec e_i)$, where $\pm \vec e_i$ are
the six vectors which connect a vertex in the triangular lattice
to its nearest neighbors within the same sublattice. The algorithm is
similar to the one in 2 dimensions:
 First, we choose
a sequence of endpoints connecting the initial to the
final endpoint which is as short as possible. Then, we draw
all the minimal paths leading to these endpoints, and attempt to
move the line through them sequentially. If two consecutive
minimal paths have nowhere a distance
larger than 1 (measured in units of $|\vec e_i|$), we can choose a
sequence of elementary moves such that at most two bonds of the
line are not on one or the other minimal path, leading to a
barrier of order 1 between the two. If the distance is larger than
1, we proceed essentially in the same way as in 2 dimensions, i.e. we
 consider the
midway points  $(t /2, \vec x_i)$ which connect both
lines  via the shortest possible
trajectory $x_i(t/2)$ (if there are several possibilities, we choose one
at random).  For each of
these points, we find two minimal segments of length $t/2$
connecting on one side
to $(0, \vec x)$ and on the other to either $(t, \vec x_1(t))$ or
$(t, \vec x_2(t))$. Then we move the line by repeatedly moving segments
of length $t/2$, etc.

The energy of a candidate barrier is then given by
\begin{eqnarray}
 &&\langle E_c^{(3d)}(t,t^\zeta)\rangle \nonumber\\
& &\simeq \langle E_0(t,t^\zeta\rangle) \,
(1 + 2\, \left((1/2)^\theta +
(1/4)^\theta + \cdots \right)\nonumber\\
 & &= \langle E_0(t,t^\zeta)\rangle\, \left( -1 + 2/(1-(1/2)^\theta)\right)
 \nonumber\\
 & & = 12.0\ldots\, \langle E_0(t,t^\zeta)\rangle . \label{Ec3d}
\end{eqnarray}
In principle, one should add correction terms similar to those in
Eqs.\ref{meanEc} and \ref{meanEc2}. However, these corrections
are subleading with respect to  $t^\theta$ and will be neglected.

As mentioned in section \ref{u3}, we cannot rule out that the
energies of minimal segments are independent from each other.
In the worst case, where they are completely dependent,  Eq.(\ref{Ec3d})
has to be multiplied by $\sqrt{\log_2 t}$. This may result in an additional
factor proportional to $ \sqrt{\ln t}$ in the upper bound, but does not
otherwise affect any of our conclusions.
The number of independent candidate barriers increases with some
power in $t$. Since their energy distribution decays like a Gaussian, we
can take their maximum in the same way as before, and
we finally  obtain the following estimate for
the upper bound,
\begin{eqnarray}
&& E_+^{(3d)}(t,|\vec x_f - \vec x_i|) = \nonumber \\
&& = \langle E_c(|\vec x_f - \vec x_i|,t) \rangle
+\sqrt{2\ln N}{\rm var} E_c(|\vec x_f - \vec x_i|,t)
\nonumber \\
& \simeq & \left(\sqrt{\ln t} \right)t^{\theta} f_+
(|\vec x_f - \vec x_i| /  t^{\zeta})\, .
\end{eqnarray}

We have thus shown that the energy barrier encountered
by a FL moving in a $2d$ or $3d$ random medium has an upper and a lower
bound which both increase as $t^\theta$, except for
logarithmic factors. It thus follows that the
barrier itself scales as $t^\theta$, confirming
the hypothesis $\psi = \theta$.
Since the arguments are mainly based on the
exponential tails of the minimal energy distributions, it is expected
that the result $\psi = \theta$ holds also in higher dimensions.
The only requirement is that the tails in  the distributions of minimal
energies still  decay sufficiently rapidly.

\section{Barriers to far from minimal configurations}
\label{u5}

In the previous sections, we discussed energy barriers
which have to be overcome by a line
moving between minimal energy configurations. We showed that such lines
can stay in an energy interval $\langle E_{min} \rangle \pm \text{const }
t^\theta$. However, a line may initially have an
energy which is much larger.
The initial configuration of a  FL penetrating the system may
be straight and parallel to the external magnetic field. If a system
is cooled down from high temperatures,  configurations of the FL
are random walks of roughness exponent
$\zeta = 1/2$. An initial configuration with roughness exponent
$\zeta = 1$ is found for FLs driven close to
a depinning transition\cite{Deniz}.
If the temperature is low (as we always
assume in this paper), the line then relaxes
to some metastable state.
The FL will ultimately reach a configuration of
minimal energy,  only if it is not separated from it by abnormally
high barriers. We therefore show in this section that the line can
reach the minimal energy region by going
only over barriers which are not larger than order of $t^\theta$.
We  specify an algorithm for moving a line of length $t = 2^n$
{\it from any initial configuration}
to one of minimal energy. The algorithm
is similar to that presented in the previous sections, and leads
to barriers of the order of  $t^\theta$.

First, we
assume that its initial roughness is not larger than that
of minimal energy paths.
Let $\{x_n(\tau)\}$ for $\tau = 0,\cdots,t$ be the initial configuration of
the line, and $\{x_0(\tau)\}$ a minimal energy configuration  with $x_n(0) =
x_0(0)$ and $x_n(t) = x_0(t)$. We then define a sequence of
paths $\{x_m(\tau)\}$, $m = 1,\cdots,n-1$, which are
constrained to go through the points $x_n(k t/2^m)$ for $k=0,1,\cdots,2^m$
and  are composed of $2^m$ minimal segments of length $t / 2^m$.
The energy of such a segment is smaller than the energy of any other
piece of a path with larger $m$ which has the same endpoints as the
segment.
We now move the line successively through this sequence of configurations,
going from the largest to the smallest value of $m$. The configurations
$\{x_{m + 1}(\tau)\}$ and $\{x_{m}(\tau)\}$ intersect each other at the points
$\tau = kt/ 2^m$, with $k = 0,\cdots,2^m$. We therefore can move the line from
the configuration $\{x_{m + 1}(\tau)\}$ to $\{x_{m}(\tau)\}$ by
successively moving segments of length $t/2^m$. In many cases,
the segments have to overcome a loop, and then we apply the algorithm
defined previously. In contrast to the previous sections,
these loops do not separate two  minimal configurations,
but one minimal segment, and another constrained at its midpoint,
a constellation which occured also in the previous sections
as an intermediate situation. Since we restricted the roughness of the
initial configuration to less than that of minimal paths,
the size of the loops does not exceed $t^\zeta$. The number of independent
candidate barriers within a loop is therefore smaller than, or equal to,
$N_c \propto (2^m)^{\alpha'}$ from previous arguments, where
the exponent $\alpha'$ depends  on the model. The energy of each
candidate path is smaller than, or equal to, the energy $E_c(2^m)$, which was
also obtained in the previous sections. The total number of loops is
less than or equal to
$1+2+\cdots+2^{n-1} < 2^n = t$, and the energy of each candidate barrier
is  certainly overestimated if we assume that all loops are of size $t$.
We therefore find an upper bound to the barrier  which
is the maximum of $t\, t^{\alpha'} = t^{1 + \alpha'}$ candidates
chosen from a distribution $P(E_c(t))$ with
$\langle E_c(t)\rangle \propto t^\theta$, and with a  Gaussian tail.
As we saw in section \ref{u3}, such a maximum scales as
$t^\theta \sqrt{\ln t}$.
We therefore have shown that a line can move from any configuration with
roughness exponent less than $\zeta$ to a minimal energy
configuration by crossing barriers which are not larger than
order of $ t^\theta$, provided that the barriers between minimal
configurations scale also as $t^\theta$.

A similar result can be obtained for any initial configuration of the line.
To demonstrate this, let us look at the
configuration $x_t(\tau) = -\tau$, which is as far as possible from a minimal
configuration. We then define a sequence of paths $\{x_m(\tau)\}$, for
$m = t - 2, t - 4,\cdots, 0$, with $x_m(\tau) = -\tau$ for $\tau \le m$,
and connecting the points $(m,-m)$ and $(t,-m)$ by a minimal path. To go
from one configuration to the next one, the line has to overcome a
loop of size no bigger than  $ (t-m)^\theta < t^\theta$. There are
consequently proportional to $ t^{1+\alpha'}$
candidates for barrier paths of length between 2 and $t$. We certainly
find an upper bound to the barrier by assuming that all these candidates
have the length $t$, and that their energies
are taken from a (half-) Gaussian distribution of width proportional to
$ t^\theta$.
The upper bound consequently scales as $t^\theta \sqrt{\ln t}$.

\section{Multiple Barriers}
\label{u6}

So far, we tacitly assumed that the activation barrier is given by the
difference of the highest energy encountered by the line and its
initial energy, just as in thermally activated chemical reactions.
This assumption, however, has no solid foundation, since the line
does not simply move over an isolated maximum, but through a random
energy landscape. In addition, it is not at all clear how results
obtained for a point-like particle in a 1-dimensional energy landscape
can be generalized to lines moving in 2- or 3-dimensional systems.
To shed some light at least on the first of these points, we study in
this section a particle in a 1-dimensional energy
landscape at low temperatures. We tilt this landscape by a small
angle to take into account the effect of an external driving force.
Using the Fokker-Planck equation, we calculate the stationary particle
current through this tilted energy landscape. We find that it is not
the difference between the maximal and initial energies,
but the difference between the maximal and
the minimal energies, which determines the activation barrier.

The Fokker-Planck equation for the probability density $P(x,t)$ of an
overdamped particle in one dimension is
\begin{equation}
{\partial P \over \partial t} = \Gamma {\partial\over\partial x}
\left( kT{\partial\over\partial x} + {\partial V(x) \over \partial x}
\right) P(x,t)\, .
\end{equation}
Here, $\Gamma$ is the inverse of the product of
the particle mass and the friction coefficient.
The potential $V(x)$ is the sum of the random potential $V_B(x)$ and
a driving term $-F x $, where $F$ is the constant driving force. Depending
on the boundary conditions, this equation has different stationary
solutions $\partial_t P = 0$. If the boundary is
an infinitely high wall at both ends of the system, we obtain the
equilibrium solution $P(x) \propto \exp(-V(x)/ kT)$, and zero current
$$j = \Gamma \left( kT{\partial\over\partial x} +
{\partial V(x) \over \partial x} \right) P(x,t) = 0\, . $$
 We instead look
for a solution where particles enter the system at one end and leave
it at the other end. This solution is most readily found by assuming
periodic boundary conditions, $P(L) = P(0)$ and $V_B(L) = V_B(0)$. This
situation corresponds to a periodic energy landscape which has been tilted,
and where each section of length
$L$ contains the same number of particles. Clearly, this leads to a
stationary flow through
the system, with particles entering a section at one end and leaving it at
the other. In the limit of small driving force $F$, the stationary
current is found by considering only terms up to linear order in $F$ (order
zero gives the equilibrium solution of the untilted system), and is given
by\cite{FokkerPlanck}
\begin{equation}
j = \Gamma F L / \left[\int_0^Le^{-V_B/kT}dx \int_0^Le^{+V_B/kT}dx\right].
\label{maxmin}
\end{equation}
In the limit $T \to 0$, the integrals are dominated by the neighborhoods
of the maximum and the minimum of the potential, leading to
$ j \propto \exp[(V_B^{max} - V_B^{min})/kT]$. This means that the particle
mobility is determined by the difference between the energy maximum and
minimum. This result is plausible since the particles explore all of the
energy landscape and  therefore also go down to the valleys and have
to come up all the way again\cite{SLV}.

\begin{figure}
\narrowtext
\centerline{\rotate[r]{\epsfysize=3.5in
\epsffile{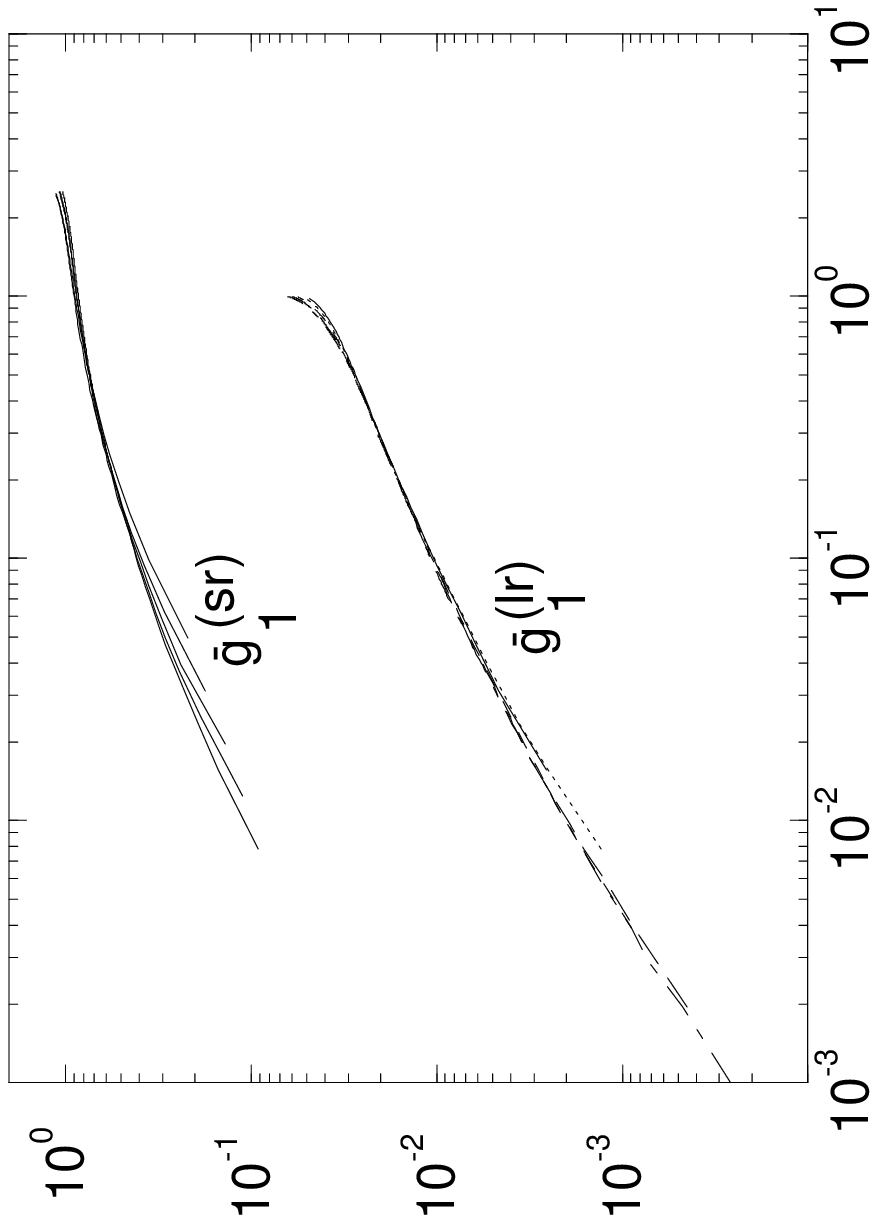}}}
\vskip -1true cm
\caption{Scaling functions $\bar g_1^{(sr)}(y)$ and $\bar g_1^{(lr)}(y)$,
similar to $ g_1^{(sr)}(y)$ in eq.(4) and $ g_1^{(lr)}(y)$ in eq.(8),
but with the barrier definition of section VI.
}
\label{upper}
\end{figure}

Generalizing the above arguments to a
line in 2 dimensions is difficult, and we did not
succeed in solving the corresponding Fokker-Planck equation
analytically. Evidently, the line can avoid
configurations with particularly high energy, which seems to justify the
assumption that the barrier is the lowest possible which
separates the initial and the final configurations.
In the light of  Eq.(\ref{maxmin}), however, we may need to define the
barrier energy as the difference between the maximum and the minimum,
instead of the difference between the maximum and the initial energy. If so,
we should add to the barrier the difference between the initial energy
and that of the absolute minimum along the trajectory of the line.
We know, however, that the distribution of minimal energies has only
exponental tails, and that therefore both types of barriers
scale in the same way. Consequently, our results do not
depend on the precise definition of the barrier.
To confirm this, we plot in Fig.\ref{upper} the scaling functions
$\bar g_1^{(sr)}$ and $\bar g_1^{(lr)}$ for the
barrier to 2-dimensional lines with either short-range or
long-range correlated randomness, determining the difference between the
maximal and minimal energy of all intermediate configurations of the line.
It can clearly be seen that the barrier energy still satisfies
Eqs.(\ref{meanE+1}) and (\ref{meanE+2}).

\section{Conclusions}
\label{u7}

In this paper we considered  various properties of the energy landscape
of one of the simplest realizations of glassy systems. We showed that,
under fairly general conditions, the energy barriers encountered by
a line descending into the region of
minimal energies, or moving within this region, scale in the same way as
the fluctuations in minimal energy. This means, in particular,
that there exist
no metastable configurations which cannot be left by going over
energy barriers smaller than, or equal to,
the  fluctuations in minimal energy.

Similar arguments  are applicable to interfaces in
random media, like domain walls in 3-dimensional
random bond and random field Ising models. When the interface
moves from an initial to a final configuration, with part of its
boundary fixed, a given boundary point moves along a line. For each
position of this boundary point, there exists a configuration of
minimal energy. The maximum of these minimal energies certainly is
a lower bound to the barrier. If the distribution of minimal energies
has no power-law tails, this lower bound scales in the same way
as the minimal energy fluctuations. An upper bound can be constructed
using a similar algorithm as for the line: Each time the interface (or
a segment of it) has to overcome a loop, we bisect it and repeatedly move
the upper and lower segment through a sequence of minimal configurations.
In this way, the interface is always composed of minimal segments, and
 it should scale in the same way as
the lower bound (except for logarithmic factors, and provided that the
lower bound energy distribution has no power-law tails).

Given these results for lines and interfaces, it is likely that they
generally hold
for elastic media with impurities, e.g. for a bunch of flux lines.
The latter situation is certainly of much more physical releance than a
single FL. Our results for a single FL,
may thus have provided a glimpse into the complexity
of the energy landscape of more complicated glassy systems.

Based on results for particles in 1-dimensional energy landscapes,
we also argue that the energy barrier should not be defined with respect
to the initial energy, but to the minimal energy along the trajectory of
the line.
It stills remains a challenge to generalize this argument to lines in 2- or
higher dimensional energy landscapes, and to find a more precise
expression for the response of the line
to a driving force, starting from the Fokker-Planck equation.

\acknowledgements
This work started in collaboration with Lev Mikheev (see Ref.\cite{MDK}).
We have also benefitted from discussions with Alan Middleton, who has
independendly discovered many interesting results pertaining to energy
barriers\cite{Middleton}. BD is supported  by the  Deutsche
Forschungsgemeinschaft (DFG) under Contract No.~Dr 300/1-1. MK
acknowledges support from NSF grant number DMR-93-03667.

\end{multicols}

\begin{references}{}
\bibitem{KardarRev}
M. Kardar, {\it Lectures on Directed Paths in Random Media},  Les
Houches Summer School on Fluctuating Geometries in Statistical
Mechanics
and Field Theory, August 1994 (to be published; see
cond-mat/9411022).

\bibitem{glass1}
K. H. Fischer and J. A. Hertz, {\it Spin glasses}, Cambridge
University Press, Cambridge (1991).

\bibitem{glass2}
J. A. Mydosh, {\it Spin glasses: an experimental introduction},
Taylor \& Francis, London (1993).

\bibitem{expreview}
G. Blatter, M. V. Feigel'man, V. B. Geshkenbein, A. I. Larkin, and V.
M. Vinokur,
Rev.\ Mod.\ Phys.\ {\bf 66}, 1125 (1994).

\bibitem{BoseGlass}
D. R. Nelson and V. M. Vinokur, Phys.\ Rev.\ B\ {\bf 48}, 13060
(1993).

\bibitem{KimAnderson}
P. W. Anderson and Y. B. Kim, Rev.\ Mod.\ Phys.\ {\bf 36}, 39 (1964).

\bibitem{FisherFisherHuse}
D. S. Fisher, M. P. A. Fisher, and D. A. Huse, Phys.\ Rev.\ B\ {\bf
43}, 130 (1991).

\bibitem{JoffeVinokur} L. Ioffe and V. M. Vinokur, J.\ Phys.\ C\ {\bf
20}, 6149 (1987).

\bibitem{HuseHenley}
D. A. Huse and C. L. Henley, Phys.\ Rev.\ Lett.\ {\bf 54}, 2708
(1985).

\bibitem{MDK}
L. V. Mikheev, B. Drossel, and M. Kardar, Phys. Rev. Lett. {\bf 75}, 1170
(1995).

\bibitem{barrier3d} B. Drossel, J. Stat. Phys., in press (1995).

\bibitem{HuseHenleyFisher}
D. A. Huse, C. L. Henley, and D. S. Fisher, Phys. Rev. Lett. {\bf 55}, 2924
(1985).

\bibitem{HwaFisher}
T. Hwa and D. S. Fisher, Phys.\ Rev.\ B\ {\bf 49}, 3136 (1994).

\bibitem{Mikheev} L. V. Mikheev, preprint (1994).

\bibitem{Galambos} J. Galambos, {\it The Asymptotic Theory of Extreme Order
Statistics}  (John Wiley \& Sons, New York, 1978).

\bibitem{MKJAP}
M. Kardar, J. Appl. Phys. {\bf 61}, 3601 (1987).

\bibitem{correlated} E. Medina, T. Hwa, M. Kardar, and Y.-C. Zhang,
Phys. Rev. A {\bf39}, 3053 (1989).

\bibitem{AmarFamily}
J. G. Amar and F. Family, Phys. Rev. A {\bf 41}, 3399 (1990).

\bibitem{KimBrayMoore}
J. M. Kim, A. J. Bray, and M. A. Moore, Phys. Rev. A {\bf 44}, 2345 (1991).

\bibitem{HalpinHealy} T. Halpin-Healy, Phys. Rev. A {\bf 44}, R3415 (1991).

\bibitem{randomresistor} V. Ambegaokar,
B. I. Halperin, and J. S. Langer, Phys. Rev. B {\bf 4}, 2612 (1971).

\bibitem{Deniz} M. D. Ertas and M. Kardar, Phys. Rev. Lett. {\bf 73}, 1703
(1994).

\bibitem{FokkerPlanck} H. Risken,
{\it The Fokker-Planck equation}, Springer-Verlag, Heidelberg (1984),
Eq.(11.45).

\bibitem{SLV} More general discussions of the motion of a particle in a
one-dimensional energy landscape can be found in
S. Scheidl, Z. Phys. B {\bf 97}, 345 (1995); and in
P. Le Doussal and V. M. Vinokur, to be published.

\bibitem{Middleton}
A. Middleton, unpublished.


\end{references}
\end{document}